\documentstyle[sprocl]{article}

\def\al{\alpha}

\def\be{\begin{equation}}
\def\ee{\end{equation}}
\def\bea{\begin{eqnarray}}
\def\eea{\end{eqnarray}}

\begin{document}

\title{The Holographic Principle and the Renormalization Group}

\author{Enrique Alvarez\footnote{
Contribution to the {\em Encuentros Relativistas ERE-98}
 (Luis Bel's Festschrift)} and C\'esar G\'omez }

\address{Instituto de F\'{\i}sica Te\'orica\\ 29049 Madrid, Spain}


\maketitle\abstracts{
We prove a c-theorem for holographic theories.}

\section{A Proposal for a Holographic c-Function}

In \cite{zamo} Zamolodchikov proved that for local and unitary two-dimensional field theories
there exists a function of the couplings , hereinafter  called the c-function, $c(g_i)$, such that
\be
- \beta_i \partial^i c \leq 0
\ee
along the renormalization group flow. For fixed points of the flow, the c-function reduces to the central extension
of the Virasoro algebra. Some generalizations of the c-function to realistic four-dimensional theories
have been suggested; let us mention in particular Cardy's proposal on $S^4$:
\be
c \equiv \int_{S^4} \sqrt{g} < T >
\ee
where $T$ is the trace of the energy-momentum tensor of the theory. There is no complete agreement as to whether
a convincing proof of the theorem exists for dimension higher than two (cf. \cite{latorre} for a recent attempt).
\par
It is not difficult \cite{ag} 
to invent a holographic definition of the central extension for N=4 super Yang-Mills
theories using the AdS/CFT map \cite{map}, especially in the light of the IR/UV connection
pointed out by Susskind and Witten in \cite{sw}.
In order to understand this construction properly, let us recall the original 't Hooft's presentation of the holographic
principle \cite{'th}, stemmning from the Bekenstein-Hawking entropy formula for black holes \cite{bh}.
Given a bounded region of instantaneous $d-1$  space $V$  of volume $vol_{d-1}(V)$, holography states
 that all the physical information on processes in $V_{d-1}$
can be codified in terms of surface variables, living on the boundary of $V$, $\partial V$. 
More precisely, the number of holographic degrees of freedom is given by:
\be
N_{d.o.f.} \sim \frac{vol_{(d-2)}(\partial V)}{G^{(d)}}
\ee
Physically this means that we have precisely one degree of freedom in each area cell of size given by the Plack length.
In spite of the fact that the Bekenstein bounds would suggest the radical approach that {\em any} physics
in $V$ can be mapped into holographic degrees of freedom in $\partial V$, the list of theories suspected to admit
holographic projection is still small, and always involves gravity. (Susskind indeed suggested  from the beginning 
that string theory
should be holographic).
\par
Let us consider 
for concreteness 
a four dimensional CFT defined on a spacetime with topology $S^3\times\mathcal{R}$ and with the natural metric
in $S^3$. Let us also introduce an ultraviolet cutoff $\delta$, and let us correspondingly divide the sphere $S^3$ into
small cells of size $\delta^3$. The number of cells is clearly of order $\frac{1}{\delta^3}$. We would like now to define the number of degrees of
freedom in terms of the central extension as
\be
N_{dof} = \frac{c}{\delta^3}
\ee
 Notice that here the parameter $c$ plays the r\^ole of the number of degrees of freedom in each cell. If the theory we are considering
is the holographic projection of some supergravity in the bulk, it is natural to rewrite this in terms of Eq (3), but with the $vol(\partial V)$ 
now replaced by a section of the bulk at $\delta = constant$, with $\delta$ being now identified with the holographic
parameter. We are thus led to the identification
\be
c\equiv lim_{\delta\rightarrow 0} \frac {\delta^3 Vol(\partial V_{\delta})}{G_5}
\ee
In the particular example of $AdS_5\times S_5$ this yields $c = \frac{R^3}{G_5}$, with $R^4 = \alpha'^{2} N g^2$\cite{map}.

\section{Renormalization Group Flow along Null Geodesics}

We shall in this section study the renormalization group  evolution of the postulated c-function ;
that is, its dependence on  the holographic variable, $\rho$.
In order to do that, the first point is to identify exactly what we understand by {\em area} (that is,
$vol(\partial V)$. Our definition clearly involves the
quotient between an area defined close to the horizon and an {\em inertial area}, so that:
\be
c(\delta)\equiv  \frac{1}{G_d} vol_{(d-2)}({\mathcal{J}}_{\delta })vol_{(d-2)}(inertial)
\ee
The meaning of the preceding formula is as follows (cf. \cite{sw}). The first term, $vol_{(d-2)}(\mathcal{J}_{\delta})$ is the volume
computed on $\mathcal{J}$ regularized with an UV cutoff $\delta$ ($R^3/\delta^3$ in the familiar example of $ AdS_5\times S_5$).
The other factor, $vol_{(d-2)}(inertial)$ stands for the equivalent volume measured by an inertial observer which does not
feel the gravitational field (that is, $\delta^3$ in $AdS_5$). The whole thing is then divided by the d-dimensional Newton's constant.
\par
Our c-function will obey a renormalization group equation without anomalous 
dimensions:
\be
\delta\frac{\partial c}{\partial\delta} + \sum_i \beta_i(g)\frac{\partial c}{\partial g_i} = 0
\ee
which means that to prove the c-theorem we just have to show that:
\be
\delta\frac{\partial c}{\partial\delta}\leq 0
\ee
\par
Now, to study the evolution in the bulk of the c-function, we need to know how this regularized definition of area evolves as we
 penetrate into the bulk. It is now only natural to identify the UV cutoff $\delta$ with the affine parameter $\hat{u}$
introduced in the previous section such that $\hat{u}\sim 0$.
\par
We would like also to argue that it is quite convenient to study the evolution of the area along null geodesics entering the bulk.
First of all,the whole set-up is conformally invariant (which is not the case for timelike geodesics).
In addition, there is a very natural definition of tranverse space there. In a Newman-Penrose orthonormal tetrad 
\footnote{We choose to present the formulas in the four dimensional case by simplicity, but it should be clear that no 
essential aspect depends on this.},
which
is a sort of complexified light cone, because in terms of a real orthonormal tetrad, $e^a$,
\bea
l^{\mu}\partial_{\mu}&\equiv & e^{+} \equiv \frac{1}{\sqrt{2}}(e^0 + e^3)\nonumber\\
n^{\mu}\partial_{\mu}&\equiv & e^{-} \equiv \frac{1}{\sqrt{2}}(e^0 -e^3)\nonumber\\
m^{\mu}\partial_{\mu}&\equiv  & e_{T}\equiv\frac{1}{\sqrt{2}}(e^1 - i  e^2)\nonumber\\
\bar{m}^{\mu}\partial_{\mu}&\equiv &\bar{e}_{T}\equiv \frac{1}{\sqrt{2}}(e^1 + i e^2),
\eea
one can easily find the optical scalars \cite{kramer} of the geodesic congruence. One has, in particular, that
\be
\rho \equiv - \nabla_{\mu} l_{\nu} m^{\nu}\bar{m}^{\mu} = -(\theta + i \omega)
\ee

where the {\em expansion}, $\theta$, is defined by $\theta\equiv \frac{1}{2}\nabla_{\al} l^{\al}$ and the {\em rotation},
$\omega$,   is a scalar which measures the 
antisymmetric part of the covariant derivative of the tangent field: $\omega^2\equiv \frac{1}{2}\omega_{\al\beta}\omega^{\al\beta}$, with
$\omega_{\al\beta}\equiv\nabla_{[\al}l_{\beta]}$. 
\par
Let us now consider a  {\em congruence } of null geodesics. This means that we have a family $x^{\mu}(u,v)$, such that
$v$ tells in which geodesic we are, and $u$ is an affine parameter of the type previously considered. The connecting vector
({\em geodesic deviation})
$Z^{\mu}\equiv x^{\mu}(u,v) -  x^{\mu}(u, v + \delta v) $ connects points on neighboring geodesics, and by construction
satisfies
\be
\pounds (l)Z^{\mu} = 0
\ee
that is, $l^{\mu}\nabla_{\mu}Z^{\al} = Z^{\mu}\nabla_{\mu}l^{\al}$. Although the molulus of the vector $Z$ is itself not conserved, 
it is not difficult to show that its projection on $l^{\mu}$ is a  constant of motion. Penrose and Rindler call {\em abreast} the congruences for 
which this projection vanishes. In this case one can show that $h = 0$, where h is defined from the projection of the
geodesic deviation vector on the Newman-Penrose tetrad:
\be
Z^{\al} = g~ l^{\al} + \zeta ~\bar{m}^{\al} + \bar{\zeta}~m^{\al} + h~ n^{\al}
\ee
Under the preceding circumstances, the triangle $(0,\zeta_1,\zeta_2)$ is contained in $\Pi$, the 2-plane 
spanned by the real and imaginary parts of $m^{\al}$.
\footnote{In the general case, it is plain that in this way we build a $d-2$-volume}. Now it can be proven \cite{penrose} that, calling
$A_2$ the area of this elementary triangle,
\be
l^{\al}\nabla_{\al} A_2 = -(\rho + \bar{\rho}) A_2  =  2 \theta~ A_2
\ee
This fact relates in a natural way areas with null geodesic congruences.
\par
Using this information we can write at once:
\be
\hat{u}\frac{d c(\hat{u})}{d\hat{u}} = (\theta \hat{u}+ d-2 )c(\hat{u})
\ee
It is worth noting at this point that $\hat{\theta}$ is finite (it corresponds to Einstein's static universe in the
standard AdS example). The divergence in $\theta$ stems from the conformal transformation necessary to go from $\hat{g}_{\al\beta}$
to $g_{\al\beta}$, to wit:
\be
\theta = \hat{\theta} + \frac{d-2}{2} \frac{N.Z}{\Omega}
\ee
 In order that inertial and $\mathcal{J}$ units be the same, it is natural to measure inertial areas in units of $\delta \equiv \frac{\hat{u}}{
(N.Z)_{\mathcal{J}}}$, where $ (N.Z)_{\mathcal{J}}$ represents the scalar product of the vector $N^{\mu}\equiv - \nabla^{\mu} \Omega$ and
$Z^{\mu}$ computed at $\hat{u} = 0$. (This is an effect similar to the usual redshift
factor).
Doing that one gets that the first derivative  vanishes to first order:
\be
\hat{u}\frac{d c(\hat{u})}{d\hat{u}} = 0
\ee
  But we can now invocate a well known theorem by Raychadhuri \cite{r}
\be
l^{\mu}\nabla_{\mu} \theta =  \omega^2 - \frac{1}{2}R_{\mu\nu}l^{\mu}l^{\nu} -  \sigma \bar{\sigma} - \theta^2
\ee
(where  the {\em shear} $\sigma\bar{\sigma} \equiv \frac{1}{2}\nabla_{[\beta}l_{\al]} \nabla^{[\beta}l^{\al]}
 - \frac{1}{4}(\nabla_{\al}l^{\al})^2)$
\par
The  Ricci term in the above equation vanishes for Einstein spaces, and the rotation must necessarily be zero
if we want the flow lines to be orthogonal to the surfaces of transitivity; that is, that there exists a family of hypersurfaces
$\Sigma$, such that $l_{\mu} = \nabla_{\mu}\Sigma$.
\par
This shows that under these conditions
\be
\hat{u}\frac{d\theta}{d\hat{u}} < 0
\ee
which is enough to prove the c-theorem in the holographic case of present interest.
\par
More details on the geometrical approach to the holographic map can be found in
our paper \cite{ag}.

\section*{Acknowledgments}
This work ~~has been partially supported by the
European Union TMR program FMRX-CT96-0012 {\sl Integrability,
  Non-perturbative Effects, and Symmetry in Quantum Field Theory} and
by the Spanish grant AEN96-1655.  The work of E.A.~has also been
supported by the European Union TMR program ERBFMRX-CT96-0090 {\sl Beyond the Standard model} 
 and  the Spanish grant  AEN96-1664. Enrique Alvarez
 is grateful to Luis Bel for accepting
him as his student certain day a long time ago.

\end{document}